\documentclass[sigplan,screen]{acmart}

\AtBeginDocument{%
  \providecommand\BibTeX{{%
    \normalfont B\kern-0.5em{\scshape i\kern-0.25em b}\kern-0.8em\TeX}}}

\usepackage[T1]{fontenc}
\usepackage[utf8]{inputenc}
\usepackage{acro}
\usepackage{adjustbox}
\usepackage{booktabs}
\usepackage{comment}
\usepackage{todonotes}
\usepackage{colortbl}

\newcolumntype{a}{>{\columncolor{gray!10!white}}c}
\newcolumntype{x}{>{\columncolor{green!10!white}}c}
\newcolumntype{y}{>{\columncolor{blue!10!white}}c}
\newcolumntype{z}{>{\columncolor{yellow!10!white}}c}
\newcolumntype{v}{>{\columncolor{red!10!white}}c}
\definecolor{OliveGreen}{rgb}{0,0.6,0}
\definecolor{ForestGreen}{RGB}{34,139,34}
\definecolor{myblue}{RGB}{37,165,203}
\definecolor{FAUblue}{rgb}{0.000, 0.2196, 0.3961}
\definecolor{myred}{RGB}{175,32,67}

\colorlet{backgroundcol}{cyan!10!white}

\usepackage{graphicx}
\usepackage{listings}
\usepackage{float}

\acmConference[SC '23]{ACM Conference}{Tuesday–Thursday, November 14–16, 2023}{Denver, CO, USA}
  
\begin{document}






\title[Characterizing the Performance of iPIC3D]{Characterizing the Performance of the Implicit Massively Parallel Particle-in-Cell iPIC3D Code}



\author{Jeremy J. Williams, Daniel Medeiros, Ivy B. Peng, and Stefano Markidis}
\affiliation{%
  \institution{Department of Computer Science, EECS, KTH Royal Institute of Technology}
  \city{Stockholm}
  \country{Sweden}}
\email{jjwil, dadm, bopeng, markidis@kth.se}

\renewcommand{\shortauthors}{Jeremy J. Williams, et al.}

\begin{abstract}
Optimizing iPIC3D, an implicit Particle-in-Cell (PIC) code,
for large-scale 3D plasma simulations is crucial for space
and astrophysical applications. This work focuses on characterizing iPIC3D’s communication efficiency through strategic measures like optimal node placement, communication and computation overlap, and load balancing. Profiling and tracing tools are employed to analyze iPIC3D’s communication efficiency and provide practical recommendations. Implementing optimized communication protocols addresses the Geospace Environmental Modeling (GEM) magnetic reconnection challenges in plasma physics with more precise simulations. This approach captures the complexities of 3D plasma simulations, particularly in magnetic reconnection, advancing space and astrophysical research.
\end{abstract}



\keywords{iPIC3D, Magnetic Reconnection, Implicit PIC, Space Weather, Performance Analysis, Profiling and Tracing}




\maketitle

\noindent \textbf{Introduction.}
The iPIC3D code is widely used massively-parallel Particle-in-Cell (PIC) for space simulations, particularly in magnetic reconnection studies. Magnetic reconnection is a phenomenon where magnetic field lines rupture and rearrange in three-dimensional space, leading to a restructuring of plasma's magnetic topology. This process converts magnetic energy into kinetic and thermal energy, accelerating charged particles and generating intense energy bursts~\cite{MARKIDIS20101509}. 

\begin{figure}[h!]
    \centering
    \includegraphics[width=0.72\linewidth]{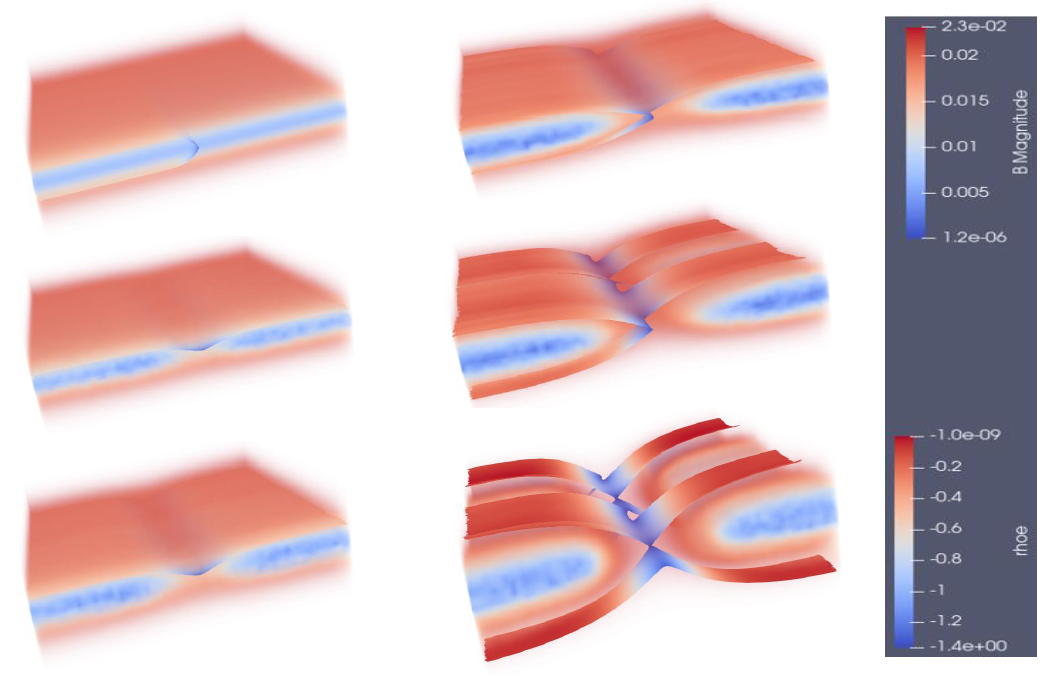}
    \caption{Evolution of streamline magnetic fields in magnetic reconnection with iPIC3D.}
    \label{fig:ipic_evolution_test}
\end{figure}

As an implicit Particle-in-Cell (iPIC) code, iPIC3D is a powerful tool for 3D plasma simulations, allowing for in-depth space exploration of plasma dynamics and complex interactions between electromagnetic fields and charged particles. It provides valuable insights into the conversion activity of magnetic energy (Figure \ref{fig:ipic_evolution_test}) and the acceleration of charged particles, shedding light on complex underlying processes in iPIC3D Plasma Simulations~\cite{MARKIDIS20101509}. 

This work aims to comprehensively analyze and enhance the performance of iPIC3D by exploring metrics like execution time, memory usage, and computational efficiency in its simulations. Advanced performance analysis tools, including profilers and tracers \cite{williams2023leveraging}, are employed to examine iPIC3D's performance characteristics, identify hotspots, and gain insights into its behavior in plasma simulations. Optimizations based on these findings seeks to improve iPIC3D's effectiveness and deepen our understanding of plasma dynamics.

\noindent \textbf{Methodology \& Experimental Setup.} 
In this work, we use \texttt{perf} as a profiler to collect hardware performance counters, focusing on cache and memory performance. We also leverage \texttt{CrayPAT} and \texttt{Apprentice2} for parallel data processing visualization on Cray architectures \cite{budiardja2018using}. We incorporate \texttt{Extrae} and \texttt{Paraver}, from the Barcelona Supercomputing Center (BSC), into our workflow for parallel performance tracing and profiling \cite{servat2013framework}. Finally, we also display our obtained results with \texttt{Darshan}, a performance monitoring tool for analyzing serial and parallel I/O workloads \cite{snyder2016modular}.

In our experiments, we analyze iPIC3D on two systems: a workstation (\texttt{Greendog}) with an i7-7820X processor (8 cores) and \texttt{Dardel}, a HPE Cray EX supercomputer with 1270 nodes, each equipped with 256GB DRAM and two AMD EPYC Zen2 2.25 GHz 64-core processors per node. Further details about these systems are provided in the poster.




\noindent \textbf{Results and Analysis.} 
iPIC3D uses parallel processing within a single node and inter-node communication for larger-scale simulations across multiple nodes. This section analyzes intra-node and inter-node activity of iPIC3D, focusing on up to 32 nodes (4096 MPI Processes) and evaluating its I/O performance.
 
We begin by assessing iPIC3D’s performance and memory system. We used \texttt{perf} on the Greendog workstation, leveraging administrative privileges (without restrictions). The impact of the iPIC3D cache size on load misses varies with cache level and workload. Larger cache sizes result in a reduction of L1 Ddcache load misses. For example, 50\% Increase Size (6 6 6 6) has 1.99\% load misses, Baseline Size (4 4 4 4) has 2.22\% load misses, and 50\% Reduction Size (2 2 2 2) has 3.79\% load misses. Similarly, for LLC load misses, 50\% Increase Size has 54.75\%, Baseline Size has 58.03\%, and 50\% Reduction Size has 47.95\%.

Next, we employed \texttt{Extrae} and \texttt{Paraver} on Dardel to analyze and trace iPIC3D's communication pattern. This analysis focused on 8 MPI ranks, with one simulation cycle (\texttt{ncycles = 1}) and all other parameters fixed, providing clear communication pattern results for a complete simulation of iPIC3D. During the initial phase to around 50\% of the simulation, the iPIC3D MPI ranks remain stable. However, ranks 1 to 7 experience a waiting period for rank 0, leading to workload imbalance and impacting efficiency in each node.

\begin{figure}[h!]
    \centering
    \includegraphics[width=1\linewidth]{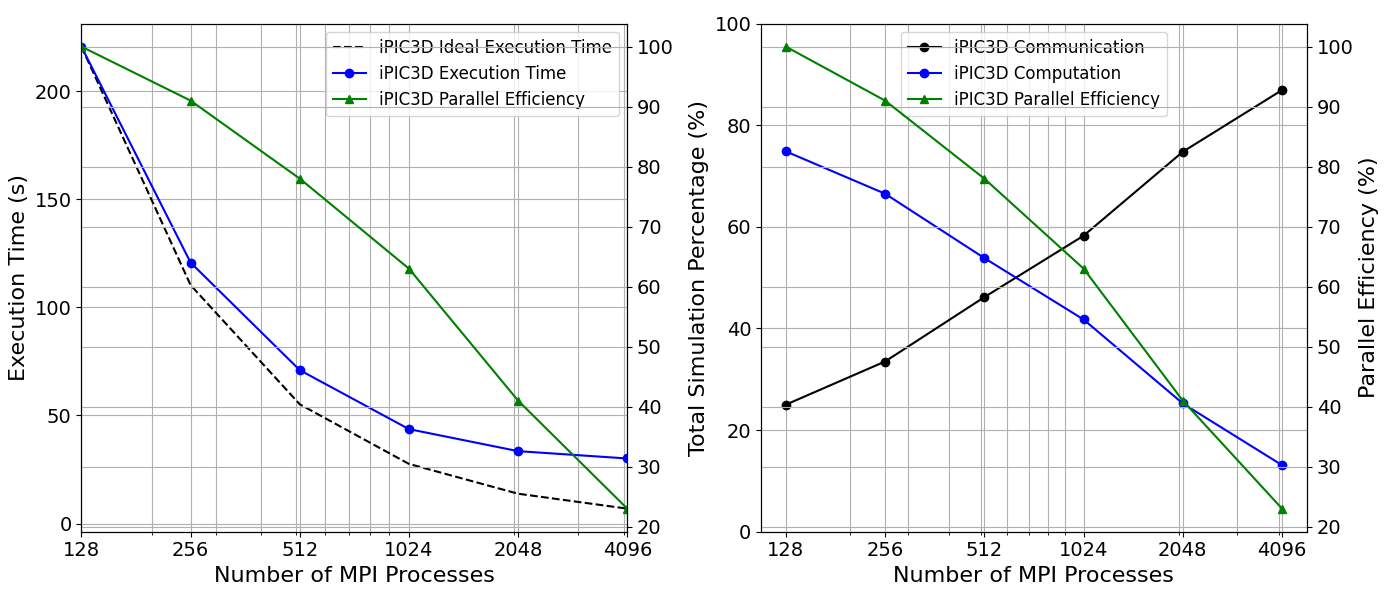}
    \caption{iPIC3D's strong scaling up to 4096 MPI processes.}
    \label{fig:craypat1}
\end{figure}

We then analyze iPIC3D parallel performance on Dardel, up to 32 nodes (4096 MPI ranks), using \texttt{CrayPAT} for instrumentation and \texttt{Apprentice2} for visualizing and exploring performance analysis data. This allows us to gain insights into the performance characteristics of multiple simulations. In the intra-node configuration (128 MPI processes), iPIC3D’s computational functions are predominantly utilized, accounting for approximately 74.84\% of the execution time. On the other hand, in the inter-node configuration (4096 MPI processes), a higher proportion of MPI functions is observed, due to increased inter-node communication demands. 
As seen in Figure \ref{fig:craypat1}, increasing MPI processes in iPIC3D results in higher communication time, reaching 86.90\% with 4096 MPI processes, while computation time decreases to 13.10\% due to overhead. Distributing computation improves overall execution time, but ideal linear scaling is not achieved, resulting in a slower execution due to communication overhead, load imbalance, and scalability limitations.



Finally, using \texttt{Darshan}, we analyze I/O in iPIC3D plasma simulations. The POSIX interface demonstrates higher bandwidth with more MPI processes, indicating better parallel performance. With 128 MPI processes, the POSIX interface achieves an I/O bandwidth of 212.28 MiB/s, while STDIO only manages 1.07 MiB/s. Increasing the number of MPI processes to 4096, the POSIX interface delivers an impressive I/O bandwidth of 4364.49 MiB/s, whereas STDIO lags behind with only 8.27 MiB/s. These results highlight how iPIC3D’s POSIX interface outperforms STDIO in I/O bandwidth, even with higher parallelism. Utilizing POSIX can significantly enhance I/O performance, especially in data transfer rates.

\noindent \textbf{Conclusions and Future Work.}
In this work, we identified communication as a critical factor and bottleneck impacting the performance in iPIC3D, particularly on large runs. Non-blocking MPI communication functions are used to mitigate this issue, however the presence of \texttt{MPI\_Waitall} can hinder execution and slow down progress. Additionally, file I/O operations (POSIX and logging) contribute to performance overhead. 

To enhance communication efficiency in iPIC3D plasma simulations, we propose several strategies. These include optimal node placement \cite{younis2008strategies}, communication and computation overlap \cite{marjanovic2010overlapping}, and load balancing \cite{zhang2005workload}. Additionally, we suggest exploring alternative algorithms and data structures to minimize overhead, drawing insights from further advanced tooling techniques \cite{williams2023leveraging}. 

\end{document}